# Single-Shot Multi-Stage Damage and Ablation of Silicon by Femtosecond Mid-infrared Laser Pulses

Kevin Werner[1,2,*], Vitaly Gruzdev[3], Noah Talisa[1], Kyle Kafka[1,4], Drake Austin[1,5], Carl M. Liebig[5], and Enam Chowdhury[6,7,1]

[1]The Ohio State University, Department of Physics, Columbus, OH 43224, USA
[2]BAE Systems, 130 Daniel Webster Hwy., MER15-1813, Merrimack, NH 03054, USA
[3]University of New Mexico, Department of Physics and Astronomy, Albuquerque, NM 87131, USA
[4]University of Rochester, Laboratory for Laser Energetics, Rochester, NY 14623, USA
[5]Air Force Research Laboratory, Materials and Manufacturing Directorate, Wright-Patterson Air Force Base, OH 45433, USA
[6]The Ohio State University, Department of Material Science and Engineering, Columbus, OH 43224, USA
[7]The Ohio State University, Department of Electrical and Computer Engineering, Columbus, OH 43224, USA

*Kevin.Werner@baesystems.com

## ABSTRACT

Although ultrafast laser materials processing has advanced at a breakneck pace over the last two decades, most applications have been developed with laser pulses at near-IR or visible wavelengths. Recent progress in mid-infrared (MIR) femtosecond laser source development may create novel capabilities for material processing. This is because, at high intensities required for such processing, wavelength tuning to longer wavelengths opens the pathway to a special regime of laser-solid interactions. Under these conditions, due to the $\lambda^2$ scaling, the ponderomotive energy of laser-driven electrons may significantly exceed photon energy, band gap and electron affinity and can dominantly drive absorption, resulting in a paradigm shift in the traditional concepts of ultrafast laser-solid interactions. Irreversible high-intensity ultrafast MIR laser-solid interactions are of primary interest in this connection, but they have not been systematically studied so far. To address this fundamental gap, we performed a detailed experimental investigation of high-intensity ultrafast modifications of silicon by single femtosecond MIR pulses ($\lambda$= 2.7 – 4.2 µm). Ultrafast melting, interaction with silicon-oxide surface layer, and ablation of the oxide and crystal surfaces were *ex-situ* characterized by scanning electron, atomic-force, and transmission electron microscopy combined with focused ion-beam milling, electron diffractometry, and µ-Raman spectroscopy. Laser induced damage and ablation (LIDA) thresholds were measured as functions of laser wavelength. The traditional theoretical models did not reproduce the wavelength scaling of the damage thresholds. To address the disagreement, we discuss possible novel pathways of energy deposition driven by the ponderomotive energy and field effects characteristic of the MIR wavelength regime.

## Introduction

The field of Intense mid-infrared (MIR) laser matter interaction has recently gained wide ranging interests due to novel phenomena like photon acceleration in metasurfaces[1] to generation of attosecond pulses reaching carbon K-edge[2], and demonstration of megafilamentation in atmosphere[3], which may open doors to many exciting new applications[4] and help probe novel phenomena[5]. Intentional MIR fs-LID can be used for the sub-surface micromachining of multi-layer materials, such as the fabrication of buried silicon waveguides operating in the telecommunications band[6]. Motivated by promise of such science and technology advances, there are several major efforts under way for the development of large MIR laser facilities such as the Extreme Light Infrastructure (ELI ALPS)[7] in Europe or BESTIA (Brookhaven Experimental Supra-Terawatt Infrared at Brookhaven Accelerator Test Facility)[8]. The ability to accurately predict MIR fs-LIDT is crucial to the design and development of new sources[9] in this wavelength regime. However, femtosecond laser damage and material modifications studies performed so far has been mostly limited to using near-infrared (NIR) wavelengths (near 780 nm, see supplemental material for references). Thus at present, a significant gap remains in theoretical understanding and experimental data and results for fs-LIDA MIR wavelengths[10, 11], which substantially slows down the fundamental and applied research in this area. In this report, we present a systematic study of irreversible ultrafast MIR laser interactions with the single crystal silicon surface (Fig. 1).

Under the single frequency approximation, the ponderomotive energy of an oscillating conduction-band electron:

$$W_\mathrm{P} = \frac{e^2 E_0^2}{4 m_\mathrm{CB} \omega^2} \quad (1)$$

is expressed via effective electron mass $m_{CB}$; carrier laser frequency $\omega$, and peak electric field $E_0$ of the driving laser pulse. Equation (1) assumes the amplitude of the laser-driven momentum oscillations is $eE_0/\omega$. Due to $\lambda^2$ scaling of the energy of Eq. (1), the ponderomotive energy of high-intensity MIR pulses can exceed the energy of a single photon and introduce novel effects into the process of absorption of laser-pulse energy. To qualitatively identify them, we note that, in a typical semiconductor, ultrafast absorption is coupled to three types of electronic excitations: inter-band transitions from valence to conduction band separated by a band gap of few eV; intra-band transitions within the same energy band; and transitions from the valence band to defect levels or from defect levels to the conduction band[12]. For interaction of good-quality crystals with sub-200-fs laser pulses, contributions of the defect-to-band and band-to-defect transitions can be neglected since they involve electron-defect interactions that require more time than the pulse duration[13, 14].

Intra-band excitation/free-carrier absorption[15] and the inter-band transitions corresponding to the multiphoton regime of the Keldysh photoionization model[16] both correspond to a regime of quantized variations of electron energy due to absorption of single or multiple laser photons; a *photon-driven* absorption regime. Here, tunneling ionization is negligible[17, 18]. In this regime, $W_P \ll \hbar\omega$.

In the opposite extreme, we can consider the case of photon energy small enough for continuous variations of electron energy (i.e. about 0.04 eV corresponding to $\lambda \geq 25$ µm; the room temperature electron energy[19]). Within this *field-driven* absorption regime, tunneling ionization is the dominant source of inter-band excitation[16]. Specific features of tunneling ionization[20] suggest that intra-band absorption of the ponderomotive energy dominates energy transfer[19].

Therefore, a transition from the quantized photon-driven absorption to the continuous field-driven one can be expected in the wavelength range from 1 to 25 µm (corresponding photon energy range is 1 eV – 0.04 eV) for typical semiconductors. Studies of wavelength scaling of laser-induced damage or ablation threshold can substantially assist in a search for the fundamental transition between the two ultimate absorption regimes.

Experimental data and theoretical results[11, 17, 18] suggest an increasing LIDT with increasing wavelength in the *photon-driven* regime. An ultimate threshold of the field-driven breakdown is that by DC electric field in a semiconductor[21], but it is orders of magnitude lower than peak time-dependent field of femtosecond LID thresholds[11, 18, 22-28]. Therefore, it is reasonable to a transition between the photon-driven and field-driven regimes which is expected to produce a maximum LIDT for some wavelength in the range 1 to 25 µm.

In this work, multiple stages of damage and ablation of single-crystal silicon by MIR femtosecond pulses are reported, and scaling of single-pulse LID thresholds of each individual stage as a function of laser wavelength is measured experimentally, demonstrating a maximum threshold in the explored MIR part of the optical spectrum. These observations show a significant qualitative difference between the strong-field ultrafast laser-solid interactions at MIR wavelength and those at ultraviolet, visible, and near-infrared wavelengths, for which absorption is assumed to be photon-driven[11, 18, 22-24, 27, 29-33]. Estimations of the Keldysh parameter demonstrate that a transition from multiphoton to tunneling regime of the photoionization happens at shorter wavelengths than the position of the damage-threshold maximum. Therefore, the maximum of the wavelength scaling of ablation threshold is reached under domination of the tunneling ionization and cannot be attributed to the transition between the two photoionization regimes. Two traditional models of laser-induced damage based on the Gamaly approximation[29] and the Keldysh photoionization theory[16] fail even to qualitatively reproduce the wavelength scaling of the damage/ablation threshold. Qualitative analysis of involved mechanisms suggests attributing of the experimental data to the transition from the photon-driven to field-driven regimes of absorption that is substantially influenced by a competition between different inter-band excitation paths and laser-induced modification of band structure of silicon[34].

## Results

Single-pulse MIR fs LID on single-crystal silicon (see Methods) was characterized ex-situ by scanning electron microscopy (SEM), atomic force microscopy (AFM), cross-sectional transmission electron microscopy (TEM), and $\mu$-Raman spectroscopy (Figs. 1-4).

Figure 2 shows representative SEM and AFM images of a damage spot. Three characteristic areas of surface modification were observed, and Figure 2a shows the approximate local fluence within each of the areas. The SEM reveals a central ablation spot, surrounded by a distinct crater rim (area 4 in Figure 2a). Area 3 in Figure 2a is enclosed by the middle ring and shows a similar contrast to the central crater. The third ring, barely visible, encloses area 2 (Figure 2a), and exhibits a change in contrast compared to the other two areas. Area 1 in Figure 2a has no outer boundary, and represents the undamaged silicon substrate with no permanent changes detected via SEM or AFM ex-situ. In the AFM spatially resolved depth profile of the same damage site (Figure 2b and 2c), the same 3 rings are all visible, each with the same dimensions as in the SEM images (Figure 2a). Profiling of the AFM images (Fig. 2c) shows that area 2 is slightly raised over areas 3 and 1. Extra SEM images (see Supplementary Materials) suggest that the chaotic nature and nano-roughness within the shorter wavelength damage site is far more pronounced compared that of the longer wavelength, for a fixed fluence.

Fig. 3 depicts representative sub-surface structure of the four areas of the damage spot shown in Fig. 2. The topmost layer of the sample shown with a white color has amorphous structure as identified by electron diffraction (Figure 3f). Deeper layers of darker colors have crystalline structure (Figure 3e). Within the resolution limit of the imaging device no polycrystalline areas could be found. Figure 3d images the sub-surface layers of the innermost ring (region 4 in Fig. 2), where ablation formed the crater. A relatively non-uniform 20-25 nm thick amorphous layer is observed in the middle ring (region 3 in Fig. 2). The area bound by the barely visible outermost ring has a relatively thick (50 nm), surprisingly uniform amorphous layer. In each image of Figure 3, a very thin (1-2 nm thick) native oxide layer is visible on the sample surface in the form of a slight contrast change. Extra TEM data (see supplementary Materials) suggest that the middle ring and the onset of very thick amorphization do not occur in the same location. The middle ring boundary has a lower fluence threshold than the crystalline/amorphous boundary. The crystalline/amorphous boundary occurs ~2 um closer to the center than the outer ring.

Spatially resolved micro-Raman spectroscopy of a typical damage site (Fig. 4) was performed at the 520 cm$^{-1}$ Raman peak associated with the crystalline phase (Fig. 4a) and the 480 cm$^{-1}$ Raman peak attributed to amorphous structure (Fig. 4c). The micro-Raman images match up with a SEM image of the damage spot (Fig. 4b) to properly attribute the crystalline and amorphous contents to the specific parts of the damage spot. Red dotted lines are aligned to certain transitions in either the Raman mapping or the SEM. Line (a) shows the transition from the highly amorphous region to the un-damaged sample Line (b) is aligned to the middle ring of the SEM image within the highly amorphous region. On the Raman mapping, this ring is not distinguished, indicating that there is no significant change in crystallinity below this ring. Line (c) is aligned to the transition between highly amorphous and highly crystalline signals on the Raman mapping. This transition occurs several um closer to the center than the middle ring. Line (d) shows that the areas both inside and outside the rim show a strong crystalline signal. Overall, the Raman mapping is consistent with the TEM images of Fig. 3.

Figure 5a illustrates the approach to measurement of LID thresholds for the three most probable damage effects under consideration: phase explosion (center of a LID spot), ablation, and ultrafast melting (most outer part of the spot). The diameter of each ring, on each damage site, for each wavelength, was measured from SEM images. The data were then plotted as dependence of diameter squared on logarithm of peak fluence to determine LIDT (Figure 5a) for each area. Scaling of the damage thresholds with wavelength are plotted in Figure 5b. Damage thresholds remain constant or rise slightly with increasing wavelength, except for the longest wavelength where a drop in damage thresholds occurs for every regime. A comparison of the experimental wavelength scaling to the predictions of the Gamaly[29] and Keldysh[16] models is shown in Fig. 5b (see details of the simulations in Materials and Methods and in Supplementary Materials). The Gamaly model[29] qualitatively agrees with the strong ablation experimental data and even captures the overall trend for much of the data at shorter wavelengths for indirect band gap, but fails to capture the reduction in LIDT for the longest wavelength. The simulations with the Keldysh photoionization model[16] greatly overestimate the experimental LIDTs and do not even qualitatively reproduce the experimentally observed trend.

## Discussion

SEM (Fig. 2), AFM (Fig. 2), and TEM (fig. 3) images as well as the Raman spectra mapping (fig. 4) suggest three stages of surface MIR laser damage each characterized by a well-defined threshold (Fig. 5A). Formation of amorphous silicon in the outermost region indicates that the velocity of the cooling front exceeds the critical speed of amorphization in silicon (12-25 m/s)[22]. This region raises slightly above the sample surface and exhibits significant uniformity of the amorphous-layer thickness attributed to the Gaussian space distribution of fluence. These features, along with the measured threshold, are consistent with those of the ultrafast melting at NIR wavelengths[35, 36], except the reduced thickness (30 nm for NIR versus 50 nm for MIR) attributed to strong linear absorption of the NIR photons. Therefore, it is reasonable to assume that the ultrafast melting induced at MIR wavelengths is driven by the mechanism established for the NIR wavelengths: bond softening by increase of conduction-band electron density to approximately 10% of the valence band electron density[36-40], about $10^{21}$ cm$^{-3}$. It is believed the energy of the excited electrons does not substantially affect that process[38, 39].

It is remarkable that the plasma critical density ($10^{19} – 10^{20}$ cm$^{-3}$) frequently considered as a criterion to determine LID threshold[17, 22, 31, 33, 41] is well below the electron density required to initiate the ultrafast melting. Moreover, during the first 100-200 fs in silicon, experiments have shown the reflectivity change remains less than 10% even at fluence 1.5 times the melting threshold[42]. The weak variations of refractive index during a 200-fs laser pulse[10] signals that the laser-generated electron-hole plasma is substantially transparent. This could occur, for example, if the plasma is far from equilibrium[43, 44]. Therefore, a reasonable criterion for LID threshold by the ultrashort laser pulses should consider the electron density required for ultrafast melting rather than the critical plasma density.

The area bound by the middle circle demonstrates the features of associated with the ultrafast melting and amorphization, but several nm of material are removed through spallative ablation. A similar phenomena was observed at MIR wavelengths on Ge[10] and at NIR wavelengths on Si[35]. However, at NIR wavelengths, typically only the top oxide layer is removed in this zone[35]. In contrast, MIR wavelengths show removal of both the oxide layer and some underlying material.

Both the cross-sectional TEM (Figure 3) and the µ-Raman mapping (Figure 4) show a transition from amorphous to crystalline material several microns inside the middle ring. For comparison, at NIR wavelengths, a strong amorphous phase persists even partially inside the crater rim[35]. Furthermore, the single-crystal recrystallization revealed by TEM (Fig. 3) also differs from the NIR case, where polycrystalline phase was observed in the form of hillock formation within the crater rim[35]. The difference between the amorphization by MIR and by NIR pulses[35, 45, 46] is partly justified by the fact that the MIR experiments were performed on a (100) surface while the NIR experiments were reported for a (111) crystalline surface. It has been shown that a (111) crystal surface is amorphized more readily after laser-induced melting than the (100) surface[45-47], while the transition from amorphization to crystallization occurs at a lower threshold fluence on a (100) surface compared to (111). Moreover, (111) surfaces were found to recrystallize into a polycrystalline phase at higher fluences when compared to (100) surfaces.

The structure of initial surface (including native oxide layer) and the specific band structure of silicon[34] are favorable for a variety of laser-stimulated process that can contribute to the surface modification. The most rigorous approach to interpretation of the reported above data should consider ab-initio simulations of the ultrafast strong-field MIR interactions[48]. However, available ab initio simulations miss a proper incorporation of electron-particle collisions and can be technically done over 10-20 femtoseconds of the laser-solid interaction[49]. Other numerical models[16, 26, 29, 33, 38, 39, 41, 50] are not valid for analysis of the high-intensity MIR laser-silicon interactions because they use the approximations that do not fit the indirect-gap band stricture of silicon. For these reasons, we limit our discussion below to qualitative analysis of the most relevant effects supported by estimations of the major interaction parameters.

The specific damage morphology discussed above (Fig. 2 and 3) suggests negligible influence of the native oxide layer on energy deposition in spite of its significant absorption around 2.6 – 2.8 µm and 4.4 µm[51, 52]. A primary ablation of the layer would be expected at fluence close to the ablation threshold if the oxide layer appreciably contributed to the energy deposition. However, this is not the case, and the morphology of the ablated spot (Figs. 2 and 3) suggests that the surface layer was blown up from bottom at all the tested wavelengths. A reasonable interpretation of that specific damage morphology should consider a dominant absorption of laser-pulse energy by the crystalline silicon.

The major processes of energy deposition in crystalline silicon are influenced by the specific structure of indirect-gap energy bands[34]. Since amplitude of the laser-driven variations of electron momentum $eE_0/\omega$ is 3 - 5 times larger than the half-width of the first Brillouin zone at threshold fluence, contributions to the inter-band transitions may simultaneously occur in the vicinity of all the special points (Γ, X, and L[34]) of the silicon band structure. In spite of complicated nature of those processes some substantial physics of the ultrafast electron excitation and associated absorption can be delivered by simple estimations of some characteristic parameters for the energy gaps and effective masses at those points (Figure 6a)[34].

A dominating regime of the laser-driven direct inter-band electron excitation (transition paths 1, 2, and 3 in Fig. 6a) can be identified by evaluation of the Keldysh adiabaticity parameter[16]:

$$\gamma = \frac{\omega\sqrt{m\Delta}}{eE} \quad (2)$$

where $m$ is a proper value of reduced effective electron-hole mass, and $\Delta$ is an energy gap between involved valence and conduction bands at the characteristic band-structure points. Estimations for the Γ point (effective electron mass is 0.188 of free-electron mass $m_{e0}$; direct band gap is 3.2 eV[34]) are summarized in Table 1. At all wavelengths, the values of the Keldysh parameter well below 1.0. Therefore, the tunneling regime dominates for the direct inter-band transitions at the Γ and X points (see Supplementary Materials). Since parameters of the L valley support the multiphoton regime of the inter-band transitions, which rate is several orders of magnitude lower than the rate of the tunneling transitions (see Supplementary Materials), an appreciable contribution to the direct inter-band promotion of electrons is expected from the Γ and X points of the band structure.

The specific indirect-gap band structure of silicon also favors indirect inter-band transitions (Figure 6a). They can be considered as multiphoton excitation of a valence electron to a virtual state V1 (for the Γ-to-L transition) or V2 (for the Γ-to-X transition) followed by phonon interaction with the excited electron to promote it to either L or X valley of the conduction band[12, 19, 34]. Rates of the indirect transitions around the Γ point (see Supplementary Materials and Methods) can be estimated using the lifetime $\delta t_i$ ($i$=V1 or V2) of the virtual states from the uncertainty relation[20]:

$$\delta t_i \cdot \delta W_i \geq \hbar \quad (3)$$

by assuming the energy uncertainty $\delta W_i$ is as much as the indirect band gaps (i. e., 1.12 eV for the Γ-to-X transition and 2.0 eV for the Γ-to-L transition[34]). Ratio of the total probabilities $P_4$ and $P_5$ of the indirect transitions (arrows 4 and 5 in Figure 6a) to the probability $P_2$ of the direct inter-band excitation (arrow 2 in Figure 6a) can be evaluated by via the direct-transition rate $w_{MP2}$ by the Keldysh formula[16], multiphoton-excitation rates $w_{V1}$ and $w_{V2}$ for the direct transitions to the virtual states, and the time of electron collisions with polar phonons[12, 19] (see Methods):

$$\frac{P_5}{P_2} = \frac{w_{V2}\delta t_{V2}}{w_{MP2}\tau_{eph}}, \frac{P_4}{P_2} = \frac{w_{V1}\delta t_{V1}}{w_{MP2}\tau_{eph}}. \quad (4)$$

Estimations (see Supplementary Materials) suggest that the ratio of the excitation rates of equation (4) is at least of the order of $10^1$-$10^2$ for V1 and $10^2$-$10^3$ for V2 at wavelength 2.75 μm. The ratio of the virtual-level lifetime $\delta t_{V1}$ and $\delta t_{V2}$ (about 0.5 – 1.0 fs) to the electron-phonon collision time (30-60 fs[53, 54]) is of the order of 0.01. Therefore, the rate of the indirect Γ-to-X excitation is larger than the rate of the direct transitions around the Γ-point. However, with increase of wavelength, the ratio of the excitation rates of Eq. (4) reduces by few orders of magnitude because the tunneling regime of the photoionization exhibits very weak dependence on wavelength[16], but the rate of the multiphoton excitation to the virtual levels V1 and V2 substantially decrease with reduction of photon energy (see Supplementary Materials). Therefore, within the MIR regime, a significant contribution of the indirect Γ-to-X transitions to the overall free-carrier generation at shorter wavelengths is replaced by domination of the direct inter-band tunneling excitation at the longer wavelengths.

An attempt to evaluate the rate of the Γ-point direct inter-band electron transitions by the Keldysh formula[16] applied to three valence bands (Figure 6b and Supplementary Materials) delivers the total rate about $10^{17}$ 1/(fs cm$^3$). Therefore, the total density of the laser-generated electron-hole pairs is of the order of $10^{19}$ 1/cm$^3$ by the end of a laser pulse according to the Keldysh model. This is approximately two orders of magnitude below the density required to initiate the ultrafast melting[38, 40]. This difference can result from the contribution of the indirect inter-band transitions that are neglected in the original Keldysh model[16], but also from the simplified energy-momentum relation for the central part of the Brillouin zone utilized in the Keldysh formula[16].

Since the duration of a single oscillation cycle of conduction-band electrons (9.17 fs at 2750 nm to 13.83 fs at 4150 nm) is appreciably smaller than characteristic time of momentum de-phasing by electron-particle collisions in typical semiconductors (about 20 fs[55]) the collisions can be considered as a minor perturbation to the laser-driven oscillatory intra-band electron dynamics. This approach implies that the oscillating electrons and holes possess some ponderomotive energy of the oscillations that is retained in the free-electron sub-system due to the collisions. Evaluations of the ponderomotive energy at the Γ point (Table 1) deliver the values that substantially exceed the photon energy (Table 1). Moreover, at the ablation threshold, the ponderomotive energy exceeds direct band gaps, bandwidth of the lowest conduction bands of silicon (about 2. 7 eV), and even electron affinity of crystalline silicon (about 4.05 eV[34]). The latter fact means the conduction-band electrons pumped by the ponderomotive energy can be effectively emitted from the crystal to produce local violation of electric neutrality[56, 57] that is favorable for the spallation regime of ablation. Moreover, enhancement of the tunneling mechanism of electron photoemission with reduction of the driving laser frequency[58] may make an appreciable contribution to reduction of the ablation threshold with increase of laser wavelength. Therefore, the collision-driven retaining of the ponderomotive energy can be a more effective ultrafast mechanism of energy transfer from laser pulses to the crystal than absorption of single or multiple photons via the electron-photon-phonon collisions.

**Table 1.** Laser wavelength λ; photon energy $\hbar\omega$; peak fluence $F_{ABL}$ / $F_{MELT}$, peak intensity $I_{ABL}$ / $I_{MELT}$, and peak electric field $E_{0ABL}/E_{0MELT}$ of laser pulses at threshold of ablation / melting. For the ablation / melting thresholds, there are shown the Keldysh parameter $\gamma_{ABL}$ and $\gamma_{MELT}$ (for the direct transition from the LH valence to the conduction band at the Γ point); the Franz-Keldysh reductions of band gap $\delta\Delta_{ABL}$ and $\delta\Delta_{MELT}$ for LH valence band; ponderomotive energy $W_{pABL}$ and $W_{pMELT}$ of the conduction-band electron oscillations at the Γ point; and ratios $W_{pABL}/\hbar\omega$ and $W_{PMELT}/\hbar\omega$ of the ponderomotive energy to photon energy.

| λ [nm] | $\hbar\omega$ [eV] | $F_{ABL}$ / $F_{MELT}$ [J/cm²] | $I_{ABL}$ / $I_{MELT}$ [TW/cm²] | $E_{0ABL}/E_{0MELT}$ [V/nm] | $\gamma_{ABL}$ / $\gamma_{MELT}$ | $\delta\Delta_{ABL}$ / $\delta\Delta_{MELT}$ [eV] | $W_{pABL}$ / $W_{pMELT}$ [eV] | $W_{pABL}/\hbar\omega$ / $W_{PMELT}/\hbar\omega$ |
|---|---|---|---|---|---|---|---|---|
| **2750** | 0.45 | 0.65 / 0.29 | 2.56 / 1.13 | 2.37 / 1.57 | 0.35 / 0.53 | 1.75 / 1.33 | 2.80 / 1.23 | 6.20 / 2.73 |
| **3150** | 0.39 | 0.68 / 0.29 | 2.66 / 1.15 | 2.42 / 1.59 | 0.30 / 0.46 | 1.77 / 1.34 | 3.82 / 1.65 | 9.69 / 4.19 |
| **3750** | 0.33 | 0.87 / 0.33 | 3.40 / 1.29 | 2.73 / 1.69 | 0.23 / 0.37 | 1.92 / 1.40 | 6.93 / 2.64 | 20.92 / 7.97 |
| **4150** | 0.23 | 0.52 / 0.21 | 2.01 / 0.84 | 2.11 / 1.36 | 0.26 / 0.41 | 1.62 / 1.21 | 5.03 / 2.09 | 16.82 / 6.99 |

The high values of threshold intensity (Table 1) imply significant transient modifications to the band structure of silicon, i. e., distortions of the shape of the bands[59] and variations of the band gaps. . There are several mechanisms of those modifications. For example, a phenomenon similar to the dynamic Franz-Keldysh effect[48, 60] reduces band gap. Since the high-intensity limit of that effect[48] is like the one for dc electric fields[61], the band-gap reduction can be estimated as follows[61]:

$$\delta\Delta = \left[(eE_0)^2 \frac{\hbar^2}{m_{RE}}\right]^{\frac{1}{3}}, \quad (5)$$

where $m_{RE}$ is the effective reduced electron-hole mass in the direction parallel to electric field $E_0$. Estimations for <111> direction around the Γ point suggest very substantial reduction of the band gaps of silicon both at the ablation and at the melting thresholds. Moreover, significant wavelength dependence of that effect[60] suggests an enhancement of its contribution for the longer-wavelength of the tested MIR range.

The fundamental properties of silicon are favorable for several other effects that also modify the band gaps[21]. First, the top valence band and the lowest conduction band of silicon originate from the same hybrid atomic state[12, 19, 62], those two energy bands are coupled. That coupling enhanced by electric field of laser radiation[62-64] results in reduction of the band gaps[62-64]. However, its contribution cannot be estimated in a simple way and requires extended numerical simulations.

Second, band-gap shrinkage results from increase of the conduction-band population N according to the $N^{1/3}$ law[64-66]. It reduces the band gaps by 1-2 eV at the conduction-band electron density $10^{21}$ 1/cm³ and appreciably contributes to the overall modification of band structure[21].

Third, generation of the conduction-band electrons makes another significant contribution to the band-gap modification by the band filling effect[64-66]. That modification results from the dominating filling of the lowest-energy conduction states that favors absorption of one extra photon to promote the newly arriving electrons to higher-energy vacant states of the band[21]. However, the influence of the ultrafast band-filling effect may be smaller compared to the usual estimations[64-66] because of effective re-distribution of the electrons over almost entire conduction band due to the laser-driven oscillations[21].

Finally, the ponderomotive energy substantially contributes to the band-gap modification[16, 59, 67]. According to Eq. (1), that energy can make either positive or negative contribution to the band-structure modification depending on sign of effective mass at specific parts of the energy bands. For example (Figure 6c), the X valley of the conduction band (Figure 6a) is up-shifted due to positive effective electron mass while the Γ-point segment of the band is down-shifted because of the negative effective mass near the Γ point. Such ponderomotive-energy effects reduce bandwidth of the energy bands, remove degeneracy of the bands, and can produce very prominent modification of the band gaps.

The special feature of silicon is that all those effects simultaneously contribute to the band-structure modification by either band-gap reduction (e. g., the analog of the dynamic Franz-Keldysh effect, band-gap shrinkage, and enhancement of band coupling) or band-gap increase (band filling and ponderomotive energy) making overall dynamics of the band-structure modification very complicated. Moreover, the ultrafast dynamics driven by the femtosecond laser pulses can be appreciably influenced by quantum interference between different paths of the laser-driven electron excitations. Also, the effects of the non-parabolic, non-Kane conduction band topology may be very important. Time-dependent ab initio simulations are the most appropriate to attack that complicated problem, but the currently available approaches cannot properly address those effects [Yabana].

## Conclusions

Single-shot fs-LIDA experiments on silicon followed by cross-sectional TEM, SEM, AFM, and micro-Raman spectroscopy have revealed unique LIDA morphology attributed to multiple stages of surface LIDA. For certain well defined fluences, a "weak" spallative ablation occurs which was not observed for NIR wavelengths under similar conditions. Surface LID thresholds were determined for the three distinct stages of damage at four MIR wavelengths. These studies deliver significant insights into the involved damage mechanisms at the MIR wavelengths.

Ultrafast melting, followed by amorphization, was observed, like in NIR experiments, but it occurs well above the plasma critical density, despite the prediction of the high-collision-rate Drude model. Spallative ablation of a few layers of silicon is the only means by which oxide layer removal was observed with MIR wavelengths. The nanoroughness of the remaining material after ablation decreases with wavelength.

Qualitative analysis of the absorption mechanisms and associated electron excitations by the ultrashort laser pulses suggests attributing the maximum of the wavelength scaling of LID thresholds to simultaneous action of several effects. They include a transition from domination of indirect inter-band excitation to domination of direct inter-band excitation; laser-driven modification of energy bands; enhancement of tunneling electron emission; and the transition from photon-driven absorption regime the field-driven regime. As discussed above, the shorter (e. g., NIR) wavelengths are favorable for dominating contributions of the multiphoton absorption to the inter-band electron promotion, intra-band excitation, and electron emission. In contrast, increase of laser wavelength towards longer-wavelength part of the MIR range leads to reduction of photon energy, increase of ablation thresholds, and corresponding increase of ponderomotive energy of the oscillating conduction electrons. Respectively, the probability of all the multiphoton effects reduces, but the contributions of the field effects increases with increase of laser wavelength. That transition between the absorption regimes is associated with increase of the contribution of ponderomotive energy to the intra-band electron excitation and overall absorption. Therefore, the wavelength scaling of the melting and ablation thresholds may reflect a transition from the photon-dominated regime of absorption to the field-dominated regime. More detailed justification of this hypothesis requires extended ab initio simulations not available currently.

## Methods

**LID experiments.** The laser source was The Ohio State University (OSU) Extreme Mid-Infrared (EMIR) optical parametric amplifier (OPA)[1, 68, 69]. A schematic of the experiment is shown in Figure S8. 1-on-1 LID experiments were performed in air, with p-polarized pulses ($\lambda$=2.75, 3.15, 3.75, 4.15 µm, $\tau_{FWHM}$=200 fs) and an angle of incidence of 31 degrees. By varying the pulse energy, the peak fluence was varied from 0.25 to 2 Jcm$^{-2}$. The energy of every pulse was recorded. The beam profile of the focal spot illuminating the surface of the sample was recorded once for each wavelength, and focal spot sizes ranged from 20-25 µm FWHM (geometric mean of the horizontal and vertical spot sizes). Reported fluences are calculated after projection of the measured focal spot profile onto the sample surface (surface normal fluence). The sample used was an *n*-type un-doped single crystal Si (1 0 0)/<1 1 0> (resistivity >10$^3$ Ω-cm with intrinsic carrier concentration ~10$^{14}$ cm$^{-3}$); electric field polarization was primarily in the (1 1 1) direction. Ex-situ SEM imaging was performed on each damage site. AFM imaging was performed on a subset of damage sites. Cross sectional TEM and spatially-resolved $\mu$-Raman spectroscopy were both performed on a single damage site.

**The two traditional simulation approaches.** Keldysh simulations were performed using Python simulation code[70]. This code uses the Keldysh equations described in the supplemental materials. The minimum peak incident laser intensity required for ultrafast melting, i.e. for photoionization of 7% of the valence band electrons, was determined to be the LIDT. Once the conduction band electron density just barely reaches the threshold at the surface, ultrafast melting of the surface layer occurs. In this work, our simulations consider direct transitions from the valence band Γ valley to the conduction band with the positive effective mass. This is indicated through transition number 2 in Figure 6a. Specific material parameter inputs for silicon are summarized in Supplemental Materials. Additional details are provided in the supplemental materials.

**Estimations of the probabilities of direct and indirect transitions.** For derivation of the estimation of equation (4) of the main text, we assume that the virtual states V1 and V2 are reached via a direct transition by absorption of several laser photons. Correspondingly, the energy gap between the Γ point of the valence bands and the virtual states is evaluated as follows:

$$E_{V1} = \langle \frac{\Delta_{INDGV1}}{\hbar\omega} + 1 \rangle, E_{V2} = \langle \frac{\Delta_{INDGV2}}{\hbar\omega} + 1 \rangle, \quad (6)$$

where $\Delta_{INDGv2}$ = 1.12 eV for the Γ-to-X transition (5 in Figure 6a), and $\Delta_{INDGv1}$ = 2.00 eV for the Γ-to-L transition (4 in Figure 6a)[34]. Probability of the electron transitions from the top Γ-states of the valence bands to the virtual states per time interval $\delta t$ estimated by the uncertainty relations of equation (3) can be evaluated via the multiphoton-ionization rate $W_{V1}$ and $W_{V2}$:

$$P_{V1} = \frac{W_{V1}\delta t_{V1}}{N_{VB}}, P_{V2} = \frac{W_{V2}\delta t_{V2}}{N_{VB}} \quad (7)$$

where $N_{VB}$ is the total population of the involved valence bands. The other process involved into the indirect transitions 4 and 5 (Fig. 6) considers electron-phonon collisions[12, 19]. Probability $P_{elph}$ of this collision process per time interval $\delta t$ can be evaluated as follows[12, 19]:

$$P_{eph} = \Gamma_{eph}\delta t = \frac{\delta t}{\tau_{eph}} \quad (8)$$

where $\Gamma_{eph} = 1/\tau_{eph}$ is the rate of electron collisions with polar phonons[12, 19, 34]. The total probability of the transitions 4 and 5 can be evaluated in the first approximation as a product of the probabilities of equation (7) and equation (8) by assuming the electron excitation to the virtual states and the electron-phonon collisions are independent:

$$P_4 = P_{V1}P_{eph} = \frac{W_{V1}\delta t_{V1}^2}{N_{VB}\tau_{eph}}, P_5 = P_{V2}P_{eph} = \frac{W_{V2}\delta t_{V2}^2}{N_{VB}\tau_{eph}}. \quad (9)$$

In equation (8), the rate of the electron-phonon collisions considers all the collisions that result in variations of electron momentum, i.e., all contributions to the momentum transfer from phonons to the excited electrons of the virtual states. Characteristic time $\tau_{eph}$ of the electron-phonon momentum transfer (also referred to as momentum dephasing time) is about few tens of femtoseconds in typical semiconductors[55] depending on laser and material parameters.

To obtain the estimation of equation (4), the probability of the direct Γ-point transitions is evaluated via the inter-band photoionization rate $W_{MP2}$ as follows:

$$P_2 = \frac{W_{MP1}\delta t_i}{N_{VB}}, i = V1, V2. \quad (10)$$

The photoionization rates are estimated using the Keldysh formula[16] with corresponding corrections.

## Data availability

The data that support the findings of this study are available from the corresponding author upon reasonable request.

## Acknowledgements

This work was supported by Air Force Office of Scientific Research (AFOSR) grant (FA9550-16-1-0069) and AFOSR multidisciplinary research program of the university Research initiative (MURI) grant (FA9550-16-1-0013).

## Acknowledgements

This material is based upon work supported by the Air Force Office of Scientific Research (AFOSR) under award number FA9550-16-1-0069 and AFOSR MURI award FA9550-16-1-0013.

## Author Contributions

K.W., E.C., and N.T. conceived the experiments and built the OPA. K.W. and N.T. performed the LID experiments. D.A. performed the AFM measurements. K.K. performed the SEM measurements, and K.K. and K.W. analyzed the results. K.W. and D.A. oversaw the FIB and TEM of the samples. K.W. and E.C. analyzed the TEM results. C.M.L. performed the micro-Raman spectroscopy measurements, K.W. and E.C. analyzed the results. D.A. wrote the Keldysh photoionization code and Gamaly model code. D.A. and K.W. modified the code for use with Si. V.G. performed the theoretical analysis and calculations related to the proper interpretation of wavelength scaling results. All authors contributed to the preparation of the manuscript.

## Competing Financial Interests

The author(s) declare no competing financial interests.

# Figures

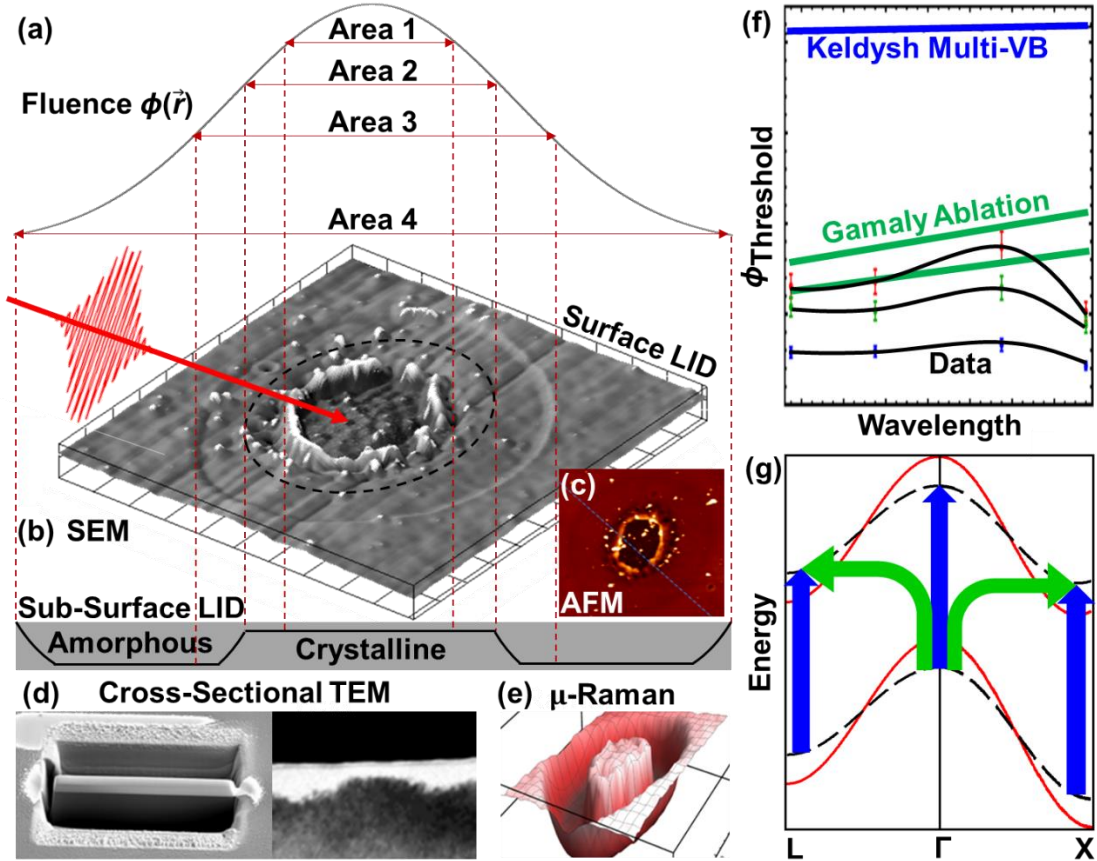

Figure 1: A schematic workflow of the research includes (a) damaging of a silicon surface by a single ultrashort laser pulse with Gaussian space distribution of fluence; detection and characterization of the four areas of a damage spot (see description in the text) by (b) SEM, (c) AFM, (d) cross-sectional TEM, and (e) μ-Raman spectroscopy; (f) measurement of wavelength scaling of the thresholds of formation of each characteristic damage-spot area followed by fitting of the scaling with the Keldysh and Gamaly models; and (g) theoretical analysis of competing direct (vertical blue arrows) and indirect (bent green arrows) inter-band electron transitions between initial (solid red) and laser-disturbed (black dashed) energy bands.

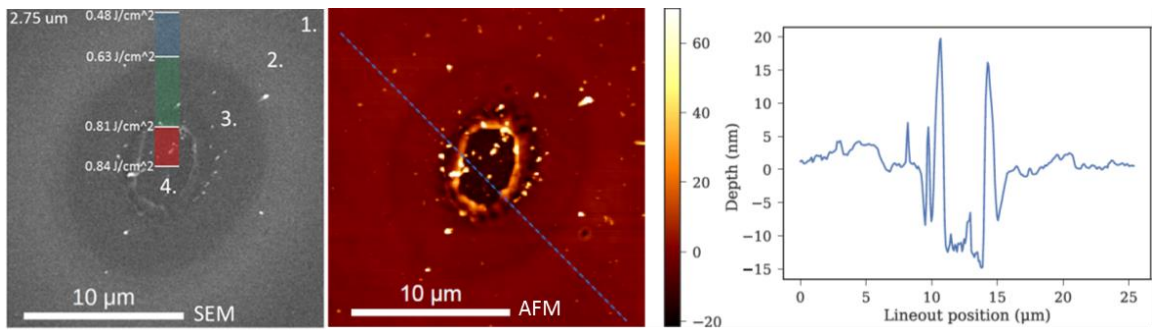

Figure 2: Side-by-side SEM (a), AFM (b), and AFM lineout (c) of the same damage spot produced at fluence of 0.84 J/cm^2 at wavelengths 2.75 μm. Areas of interest 1-4 are described in the main text. SEM overlay of panel (a) shows local fluence. AFM lineout of panel (c) is taken along the blue dotted line shown in the panel (b).

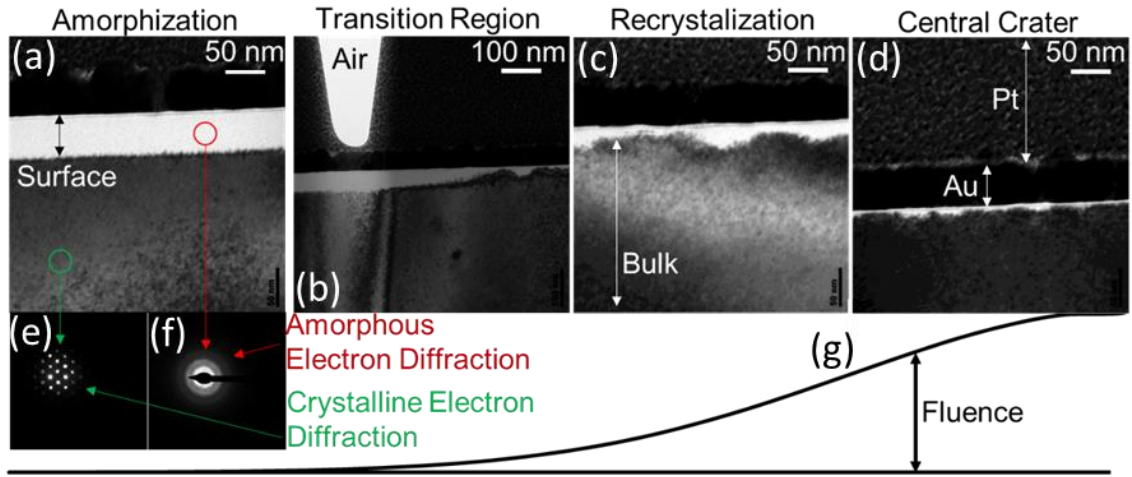

Figure 3: Cross-sectional TEM images (a)-(d) of the sub-surface damage below the same site featured in Figure 2. Four different areas of the same damage site are shown: (a) region 2 (amorphization occurs), (b) between regions 2 and 3, (c) region 3 (recrystallization occurs), and (d) region 4 (crater). The sacrificial platinum layer, protective gold layer, laser-modified surface region, and bulk silicon substrate are indicated. Electron diffraction of the darker contrast areas show a crystalline signal (e) while bright white areas indicate amorphization (f). The local incident laser fluence increases from left to right for each TEM image; i.e. images further towards the right are closer to the central crater (g). TEM was performed at the Center for Electron Microscopy and Analysis (CEMAS) at The Ohio State University.

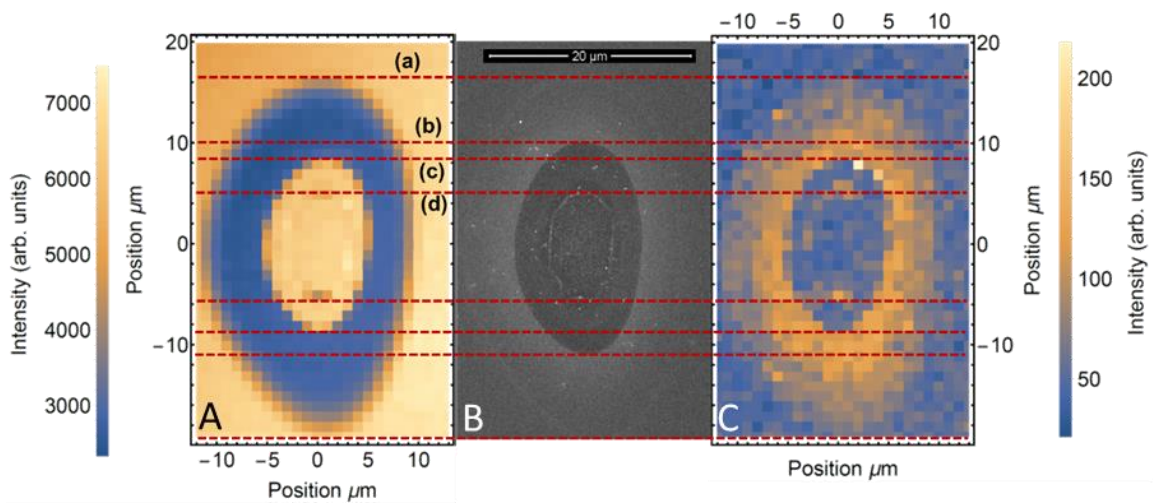

Figure 4: µ-Raman spectroscopy (A and C) of a damage spot produced at fluence of 0.87 J/cm$^2$ at wavelength 3.5 um. Peak intensity of the 520 cm$^{-1}$ signal (A) shows crystalline phase, while the 480 cm$^{-1}$ signal (C) shows amorphous phase. An SEM image of the spot (B) is shown for proper identification of different areas of the damage spot. Dotted lines (a)-(d) indicate the different areas of surface or subsurface LID discussed in the main text. µ-Raman spectroscopy measurements were performed at the US Air Force Research Laboratory in Dayton, Ohio.

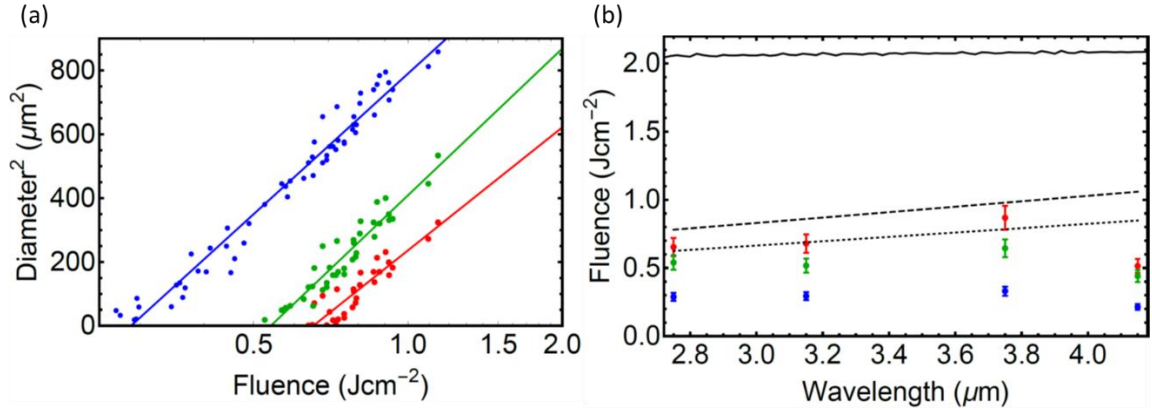

Figure 5: Scaling of area of laser-modified spot on the surface with peak laser fluence at wavelengths of 2.75 µm (A) to illustrate the approach of evaluation of threshold for phase explosion (red), ablation (green), and melting (blue) by extrapolating a linear fit towards zero area. Wavelength scaling of LID threshold as obtained from experimental data (dots with error bars), Gamaly model (dotted line – for indirect band gap; dashed line – for direct band gap), and two-band Keldysh model (solid).

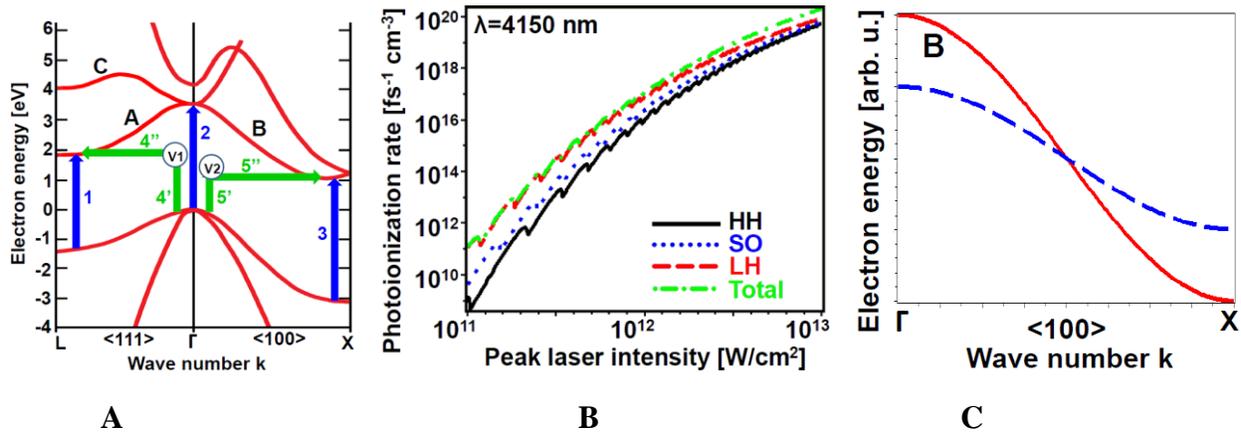

A                                                     B                                                     C

Figure 6: A sketch of competing direct (blue vertical arrows) and indirect (green bended arrows) inter-band electron transitions contributing to the nonlinear absorption and electron-hole plasma generation (a). V1 and V2 depict the two virtual states involved intot he indirect transitions. Scaling of the photoionization rate with peak laser intensity (b) at wavelength of 4150 nm to illustrate similarity of contributions of the heavy-hole (HH – black solid), light-hole (LH – red dashed), and split-off (SO – blue dotted) valence bands to the total rate of the inter-band electron excitation (green dash-dotted). A sketch of laser-induced reduction of conduction-band bandwidth along (100) direction (c) due to the modification of the initial band (red solid) by ponderomotive potential of laser-driven electron oscillations.